# Effect of Geometry on Sensitivity and Offset of AlGaN/GaN and InAlN/GaN Hall-effect Sensors

Hannah S. Alpert, Karen M. Dowling, Caitlin A. Chapin, Ananth Saran Yalamarthy, Savannah R. Benbrook, Helmut Köck, Udo Ausserlechner, and Debbie G. Senesky

*Abstract*— **The current- and voltage-scaled sensitivities and signal-to-noise ratios (SNR) (with respect to thermal noise) of various octagonal AlGaN/GaN and InAlN/GaN Hall-effect sensors were examined in this work. The effect of metal contact lengths on sensitivity and sensor offset was evaluated. Calculations that take into account the shape of the device show that devices with point-like contacts have the highest current-scaled sensitivity (68.9 V/A/T), while devices with contacts of equal length to their non-contact sides have the highest voltage-scaled sensitivity (86.9 mV/V/T). The sensitivities of the two other devices follow the predicted trends closely. All the devices have offsets less than 20 µT at low supply current operation (< 300 µA) and most remain below 35 µT at higher supply current (up to 1.2 mA). The consistent low offsets across the devices imply that the choice of Hall-effect sensor geometry should mainly depend on whether the device is current-biased or voltage-biased and the frequency at which it will operate. This work demonstrates that GaN Hall-effect sensor performance can be improved by adjusting the geometry of the Hall-effect plate specific to its function (e.g., power electronics, navigation, automotive applications).**

## I. Introduction

MAGNETIC field sensors have a wide array of applications, including position and velocity sensing in vehicles (e.g., valve positions, gear rotation speed, seatbelt buckle clamping, heading determination) and current sensing in power electronics. Devices based on the Hall effect are advantageous over other magnetic field sensing technologies because they are low-cost, easy to integrate, and linear over a wide range of magnetic fields [1].

Hall-effect sensors are typically made of silicon due to its low cost, ease of fabrication, and complimentary metal-oxide-semiconductor (CMOS) compatibility; however, silicon's narrow bandgap of 1.1 eV limits its functionality to temperatures below 200°C [2],[3]. This temperature limitation can be overcome by using a material with a wide bandgap, such as gallium nitride (GaN). In particular, heterostructures made using GaN have previously shown operation up to 1000°C [4] and radiation hardness beyond that of silicon [5],[6], making it a viable material for space applications. It has additionally become a prime material platform for power electronics monitoring due to its durable nature and potential for monolithic integration with electronics [7].

GaN heterostructures have a 2D electron gas (2DEG) that is formed when a nanometer-thick layer of unintentionally doped aluminum or indium gallium nitride (AlGaN or InGaN) is deposited on an underlying GaN buffer layer. The 2DEG, created from differences in the polarization fields of the III-nitride layers [8],[9], has a high electron mobility (1500 to 2000 cm$^2$/V·s at room temperature) [2],[5],[10]-[12], which enables high sensitivity devices. Further, 2DEG-based Hall-effect sensors have the potential for lower magnetic field offsets than silicon-based devices [13]-[15]. Junction isolated silicon-based Hall-effect sensors experience electrical nonlinearity due to the dependence of the depletion layer thickness on bias voltage [16], while 2DEG-based Hall-effect plates do not face this limitation.

In this paper, we investigate the effect of Hall plate geometry on its sensitivity and magnetic field offset. We altered the size of the Ohmic contacts to maximize the sensitivity and signal-to-noise ratio (SNR) with respect to supply current and supply voltage, and we compare our experimental results to the theoretical performance based on shape factor. While the effect of contact length on sensitivity has been discussed in previous papers [17]-[19], it has never been experimentally verified. In

This work was supported in part by the Stanford SystemX Alliance, the National Defense Science and Engineering Graduate Fellowship, and the National Science Foundation Engineering Research Center for Power Optimization of Electro-Thermal systems (POETS) with cooperative agreements EEC-1449548.

H. S. Alpert, C. A. Chapin, and D. G. Senesky* (corresponding author) are with the Department of Aeronautics and Astronautics, Stanford University, Stanford, CA 94305 United States of America (e-mail: halpert@stanford.edu; cchapin3@stanford.edu; dsenesky@stanford.edu).

K. M. Dowling and S. Benbrook are with the Department of Electrical Engineering, Stanford University, Stanford, CA 94305 United States of America (e-mail: kdow13@stanford.edu; sbenbroo@stanford.edu).

A. S. Yalamarthy is with the Department of Mechanical Engineering, Stanford University, Stanford, CA 94305 United States of America (e-mail: ananthy@stanford.edu).

U. Ausserlechner and H. Köck are with Infineon Technologies Austria AG, Villach, Austria (e-mail: udo.ausserlechner@infineon.com; helmut.koeck@infineon.com).

Section II we report on device fabrication, operation, design, and testing methodology, and in Section III we evaluate the sensitivity and offset for the four device geometries.

## II. EXPERIMENTAL METHODS

### A. Device Operation and Shape Optimization

The Hall voltage ($V_H$), measured perpendicular to both the applied current ($I$) and the external magnetic field ($B$), is defined as

$$V_H = \frac{IBr_n G_H}{qn_v t} \qquad (1)$$

where $r_n$ is the scattering factor of the material (~1 for GaN) [2],[20], $G_H$ is the shape factor, $q$ is the electronic charge, $n_v$ is the volumetric carrier density, and $t$ is the thickness of the conducting layer. For charge carriers confined in a 2D sheet, the sheet density $n_s = n_v t$. At low magnetic fields, $G_H$ depends only on the geometry of the Hall plate and the contacts; it accounts for the reduction in Hall voltage and change in linearity due to the short-circuiting effect of having finite contacts [17], [21],[22]. $G_H$ can be approximately written as a function of the effective number of squares $(L/W)_{eff}$ [23];

$$G_H \approx \frac{\left(\frac{L}{W}\right)^2_{eff}}{\sqrt{\left(\frac{L}{W}\right)^4_{eff} + \frac{\left(\frac{L}{W}\right)^2_{eff}}{2} + 4}}. \qquad (2)$$

The sensitivity of a Hall-effect device with respect to supply current ($S_i$) is proportional to $G_H$;

$$S_i = \frac{V_H}{IB} = \frac{r_n}{qn_v t} G_H. \qquad (3)$$

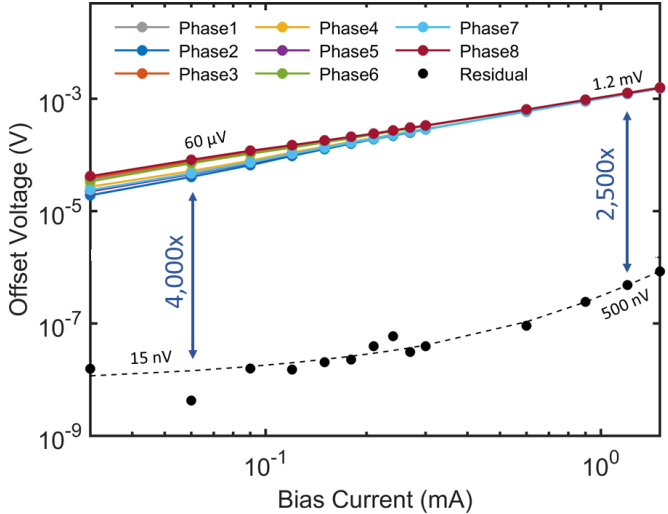

**Fig. 1.** Offset voltages of individual phases (colored lines) and the resulting offset when all phases are added (black line) as a function of bias current, for octagonal InAlN/GaN device with equal sides.

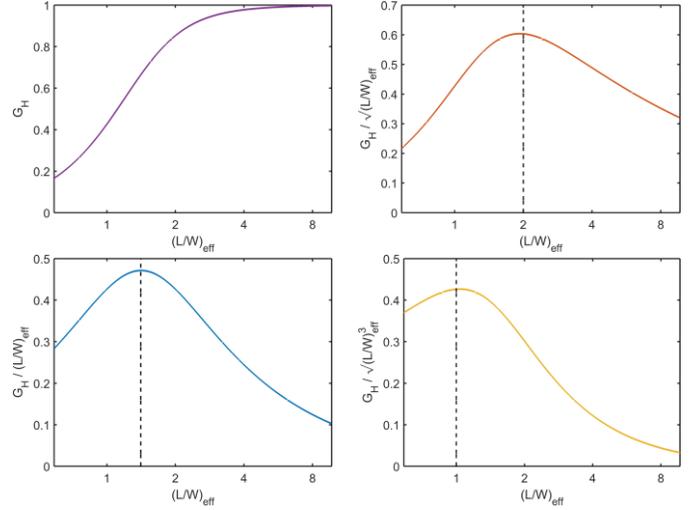

**Fig. 2.** $(L/W)_{eff}$ values that maximize important Hall-effect sensing parameters in Equations (3), (4), (7), and (8).

In addition, the sensitivity with respect to supply voltage ($S_v$) is proportional to $G_H/(L/W)_{eff}$;

$$S_v = \frac{V_H}{V_S B} = \frac{r_n G_H}{Rqn_v t} = \mu_H r_n \frac{G_H}{\left(\frac{L}{W}\right)_{eff}} \qquad (4)$$

where $V_s$ is the supply voltage, $R$ is device resistance, and $\mu_H$ is the Hall mobility of the electrons. Equations (3) and (4) show that high Hall mobility is necessary for high voltage-related sensitivity, motivating the use of the 2DEG as the sensing platform, and low sheet density is needed for a high current-related sensitivity.

In addition to high sensitivity, another desirable parameter in a Hall-effect device is low offset. The offset voltage is defined as the Hall voltage measured in the absence of a magnetic field. Offset voltages are usually caused by mechanical stress, thermal gradients, geometrical errors, defects, and other irregularities within the device [24],[25]. Implementing current-spinning has been shown to reduce the offset voltage by a factor of over 1000 [26],[27]. In this method, detailed in [24], the direction and polarity of the sourcing and sensing contacts are swapped, resulting in eight total configurations (phases) in which the Hall voltage is measured. These Hall voltages are added together, canceling out a large portion of the raw offset, shown in Fig. 1. The magnetic field offset ($B_{off}$) is calculated as

$$B_{off} = \frac{V_H}{V_s * S_v}. \qquad (5)$$

A third parameter of interest is the SNR of the device. The shape optimization described in the ensuing sections only accounts for thermal noise, as it is geometry-dependent and more significant than shot-noise and flicker noise at high frequency [11],[22]. While it was previously claimed that orthogonal switching or current spinning completely suppresses low frequency noise [28], it has more recently been shown that

TABLE I
HALL-EFFECT DEVICE CONTACT SIZES

| Device Parameter | $S_i$ | SNR/I | $S_v$ | SNR/V |
|---|---|---|---|---|
| Maximize | $G_H$ | $\dfrac{G_H}{\left(\frac{L}{W}\right)_{eff}^{1/2}}$ | $\dfrac{G_H}{\left(\frac{L}{W}\right)_{eff}}$ | $\dfrac{G_H}{\left(\frac{L}{W}\right)_{eff}^{3/2}}$ |
| $\left(\frac{L}{W}\right)_{eff}$ | $\infty$ | 2 | $\sqrt{2}$ | 1 |
| $G_H$ | 1 | 0.861 | 0.667 | 0.430 |
| $\lambda$ | 0 | 0.3 | 0.5 | 0.7 |
| Side Lengths (b = contact length) | b = 0 | 2.33 b = a | a = b | b = 2.33 a |
| Name | Point-like | Short Contacts | Equal Sides | Long Contacts |

a portion of the noise remains after spinning [13]. The thermal noise is defined as

$$V_n = \sqrt{4k_B T R \Delta f} \quad (6)$$

where $k_B$ is the Boltzmann constant, $T$ is the temperature, $\Delta f$ is the operation bandwidth, and $R$ is the device resistance across transverse contacts, which is also defined as the sheet resistance ($R_{sh}$) multiplied by $(L/W)_{eff}$. Although at low-frequency operation thermal noise is smaller than the offset voltage, it becomes significant at higher frequencies. The SNR of the device with respect to thermal noise is $V_H/V_n$, which is proportional to mobility, like $S_v$. Both SNR/I and SNR/V are directly proportional to $G_H$, while they have a dependence on $(L/W)_{eff}^{-1/2}$ and $(L/W)_{eff}^{-3/2}$ respectively, as shown in (7) and (8);

$$\frac{SNR}{I} = \frac{V_H}{V_n I} = \frac{S_i B}{V_n} = \frac{r_n G_H}{qNt}\frac{B}{\sqrt{4k_B T R \Delta f}}$$
$$= \frac{r_n G_H}{qNt}\frac{B}{\sqrt{4k_B T R_{sh}\left(\frac{L}{W}\right)_{eff}\Delta f}} \; \alpha \; \frac{G_H}{\left(\frac{L}{W}\right)_{eff}^{1/2}}, \quad (7)$$

$$\frac{SNR}{V} = \frac{V_H}{V_n V} = \frac{S_v B}{V_n} = \frac{r_n G_H \mu_H}{\left(\frac{L}{W}\right)_{eff}}\frac{B}{\sqrt{4k_B T R \Delta f}}$$
$$= \frac{r_n G_H \mu_H}{\left(\frac{L}{W}\right)_{eff}^{3/2}}\frac{B}{\sqrt{4k_B T R_{sh}\Delta f}} \; \alpha \; \frac{G_H}{\left(\frac{L}{W}\right)_{eff}^{3/2}}. \quad (8)$$

*B. Device Design*

Octagonal Hall-effect plates with four different geometries, each 100 μm across, were fabricated with metal contact lengths that maximized $S_i$, $S_v$, SNR/I, and SNR/V. Optimal $(L/W)_{eff}$ values were calculated by maximizing (3), (4), (7), and (8), shown in Fig. 2. These $(L/W)_{eff}$ values (infinity for $S_i$, 1.41 for $S_v$, 1.0 for SNR/I, and 2.0 for SNR/V) were fed back into (2) to compute the corresponding $G_H$. The ratio, $\lambda$, is defined as the length of the sides with contacts (*b*) divided by the perimeter of the full device (*a+b*, where *a* is the length of the sides without contacts). From [19] and [22], $\lambda$ was calculated for octagonal devices;

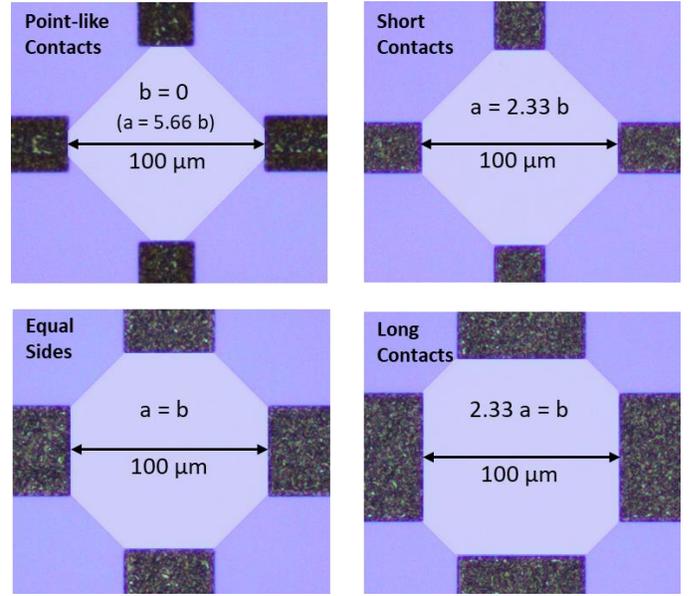

**Fig. 3.** Images of 100-μm-diameter Hall plates with various geometries, where *b* is the length of the contacts and *a* is the length of the sides without the contacts. The device with point-like contacts is optimized for $S_i$, the device with short contacts is optimized for *SNR/I*, the device with equal sides is optimized for $S_v$, and the device with long contacts is optimized for *SNR/V*. Because true point-like contacts are impossible so realize, the expression in parentheses describes the device that was fabricated.

$$G_H \approx 1 - 1.940\left(\frac{\lambda}{1+0.414\lambda}\right)^2. \quad (9)$$

The geometrical parameters involved in the optimization are summarized in Table I, and the final shapes of the Hall-effect plates are shown in Fig. 3. Because true point-like contacts are impossible to realize, the fabricated "point-like" device had a $\lambda$ of 0.165, which resulted in a $G_H$ of 0.978 and a $(L/W)_{eff}$ of 4.01, corresponding to $a = 5.66b$. The predicted percent of the SNR and sensitivity relative to the optimized shape are listed in Table II, where the values for the point-like device are based on the dimensions of the fabricated device.

TABLE II
RELATIVE DESIGN PARAMETERS OF VARIED HALL PLATE GEOMETRY

| Geometry | $S_i$ | SNR/I | $S_v$ | SNR/V |
|---|---|---|---|---|
| **Point-like** | **100%** | 80.2% | 51.7% | 28.3% |
| **Short Contacts** | 88.1% | **100%** | 91.3% | 70.7% |
| **Equal Sides** | 68.2% | 92.1% | **100%** | 92.1% |
| **Long Contacts** | 44.0% | 70.7% | 91.3% | **100%** |

## C. Fabrication

Two sets of devices were fabricated: one on an AlGaN/GaN-on-Si wafer grown by metal-organic chemical vapor deposition (MOCVD) in an Aixtron close-coupled showerhead (CCS) reactor in the Stanford Nanofabrication Facility, and the second on an InAlN/GaN-on-Si wafer purchased from NTT Advanced Technology Corporation. The AlGaN/GaN stack consists of a 1.5 µm buffer structure, a 1.2 µm GaN layer, a 1 nm AlN spacer, a 30 nm $Al_{0.25}Ga_{0.75}N$ barrier layer, and a 2 nm GaN cap. The InAlN stack consists of a 300 nm buffer structure, a 1 µm GaN layer, a 0.8 nm AlN spacer, and a 10 nm $In_{0.17}Al_{0.83}N$ barrier layer. For the AlGaN/GaN and InAlN/GaN stacks respectively, the sheet resistances at room temperature were 361 Ω/□ and 248 Ω/□ and the carrier mobilities were 1811 $cm^2$/V·s and 1143 $cm^2$/V·s. The subsequent fabrication process was followed for both sets of devices: a mesa etch was performed on the III-nitride layer, a Ti (20 nm)/Al (200 nm)/Mo (40 nm)/Au (80 nm) metal stack was deposited and annealed for 35 seconds at 850°C to form Ohmic contacts, a 7-nm-thick $Al_2O_3$ passivation layer was atomic layer deposited (ALD) to prevent oxidation, vias were etched to allow for electrical connection to the contacts, and bond metal (Ti/Au) was deposited on top. The devices were then diced and wirebonded to a printed circuit board (PCB) for testing.

## D. Experimental Testing

The devices were tested in a tunable 3D Helmholtz coil, detailed in [29]. A sourcemeter (Kiethley 2400) generated a current between two contacts across the Hall-effect plate and a multimeter (Agilent 34410A) measured the Hall voltage generated across the other two contacts. A switching matrix (U2715A) was used to alternate between the eight phases to implement current spinning [29]. During testing, the devices were placed in MuMetal® shielding cannisters to block extraneous magnetic fields; the magnetic field inside the cannisters was below 6 µT. The devices were tested with supply current ranging from 60 µA to 1.2 mA, and for sensitivity testing the applied magnetic field was ±2 mT.

TABLE III
SENSITIVITY RESULTS

|  | $S_I$ (VA$^{-1}$T$^{-1}$) |  | $S_v$ (mVV$^{-1}$T$^{-1}$) |  |
|---|---|---|---|---|
|  | Value | % Max | Value | % Max |
| *AlGaN/GaN* |  |  |  |  |
| **Point-like** | **68.85** | **100%** | 51.36 | 59.1% |
| **Short Contacts** | 61.77 | 89.7% | 72.92 | 84.0% |
| **Equal Sides** | 50.82 | 73.8% | 86.85 | 100% |
| **Long Contacts** | 33.82 | 49.1% | 80.23 | 92.4% |
| *InAlN/GaN* |  |  |  |  |
| **Point-like** | 32.18 | 100% | 35.77 | 63.7% |
| **Short Contacts** | 28.00 | 87.0% | 45.15 | 80.4% |
| **Equal Sides** | 24.20 | 75.2% | 56.13 | 100% |
| **Long Contacts** | 15.45 | 48.0% | 50.31 | 89.6% |

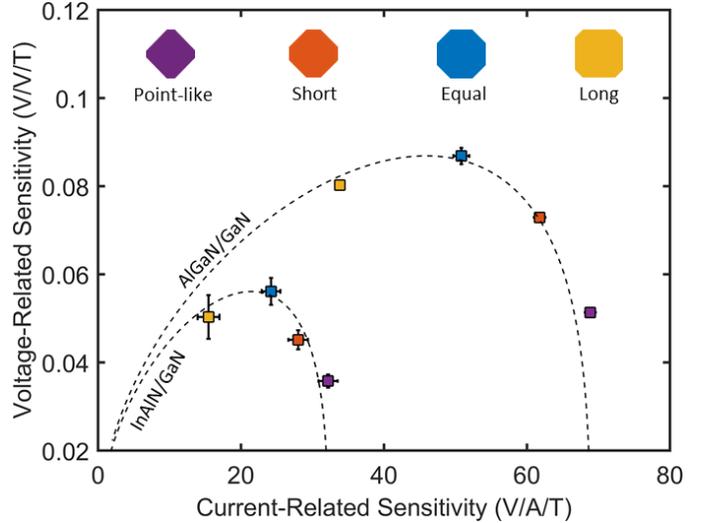

**Fig. 5.** Voltage-scaled and current-scaled sensitivity for various octagonal AlGaN/GaN and InAlN/GaN devices, where the dashed line shows the theoretical values. Both sets of devices follow the predicted trend: the devices with equal sides have the highest $S_v$ and the devices with point-like contacts have the highest $S_i$. The dotted line depicts the theoretical values, scaled according to the highest measured $S_i$ and $S_v$ for each material.

## III. RESULTS AND DISCUSSION

### A. Sensitivity

For both material platforms, the devices with the point-like contacts had the highest current-related sensitivity while the devices with equal sides had the highest voltage-related sensitivity. The measured $S_i$ and $S_v$ closely follow the predicted trends in Table II. The device sensitivities for both samples are shown in Fig. 5 and they are listed in Table III along with the percentage of the maximum value, for comparison. The AlGaN/GaN devices consistently have higher current- and voltage-related sensitivities than the InAlN/GaN devices. Since the AlGaN/GaN device has lower sheet concentration and higher mobility than InAlN/GaN, these trends hold with (3) and (4) respectively. The InAlN/GaN voltage-related sensitivities are an average of 64.7% of the AlGaN/GaN values, which closely corresponds to the ratio between their mobilities (63.1%). Similarly, the InAlN/GaN current-related sensitivities are an average of 46.3% of the AlGaN/GaN values, which matches the ratio between their carrier concentrations (45.9%).

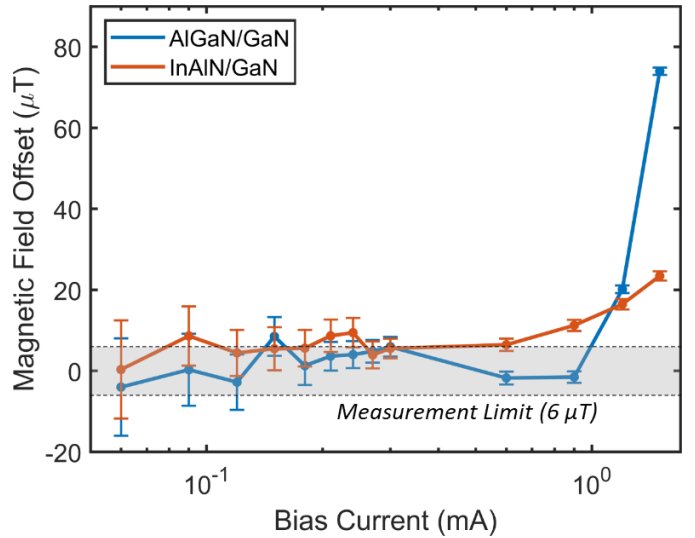

**Fig. 6.** Measured magnetic field offsets of the AlGaN/GaN and InAlN/GaN devices with equal sides. Offsets tend to be < 20 µT at low bias currents (< 300 µA) and greater at higher bias currents (up to 1.5 mA).

*B. Magnetic Field Offset*

At low bias currents (< 300 µA), the offset voltages of all the AlGaN/GaN and InAlN/GaN devices were consistently in the nanovolt range, corresponding to a magnetic field offset below 20 µT. At high biases (up to 1.2 mA), the magnetic field offsets for some of the devices remained constant below 20 µT, while some showed larger increases. The magnetic field offsets of the AlGaN/GaN and InAlN/GaN devices with equal sides are shown in Fig. 6. There is no strong correlation between Hall-effect plate geometry and offset; the variation is likely due to minor flaws during fabrication or slight differences in packaging.

## IV. CONCLUSION

We designed Hall-effect sensors to examine current- and voltage-scaled sensitivities and SNR (with respect to thermal noise), and experimentally verified how device sensitivity depends on the metal contact lengths of the Hall-effect sensor. Both the AlGaN/GaN and InAlN/GaN devices follow similar trends, confirming the validity of the shape factors over multiple material platforms. Additionally, the two material platforms confirm that increased current-related sensitivity is associated with decreased sheet density, and increased voltage-related sensitivity is associated with increased mobility. The consistent low offset of the various devices suggests that one should design or select the geometry of a Hall-effect sensor based on operating frequency and bias conditions rather than offset.